## Review Article
## Biological Value of *Centaurea damascena*: Minireview

Mohammad Jaafreh[a]; Haitham Qaralleh[b*], Muhamad O. Al-limoun[a]

[a]Biology Department, Mutah University, Mutah, Karak, 61710, Jordan; [b]Department of Medical Laboratory Sciences, Mutah University, Mutah, Karak, 61710, Jordan

*Corresponding Author: Haitham Qaralleh, [b]Department of Medical Laboratory Sciences, Mutah University, Mutah, Karak, 61710, Jordan

*Corresponding author: haitham@mutah.edu.jo

**Abstract:** The family Asteraceae include large number of *Centaurea* species which have been applied in folk medicine. One of the family Asteraceae members is the *Centaurea damascena* which authentically been tested for its antibacterial activity. The aim of the study was to discuss antibacterial activities of essential oil composition and methanolic extract of the same plant aerial part leaves. Thirty-seven components were characterized with 86% of oxygenated terpenes. The composition was dominated by 11.45% Fokienol, 8.8% thymol, 8.21% Alpha Terpineol, 7.24% Chrysanthemumic acid, 7.13% Terpinen-4-ol and 6.59% Borneol with a high degree of polymorphism in the occurrence of these compounds as compared with the different species of *centaurea*.. Free radical scavenging capacity of the *C. damascna* methanol extract was calculated by DPPH and FRAP test. DPPH radicals were scavenged with an IC50 value of 17.08 µg /ml. Antioxidant capacities obtained by the FRAP was 51.9 and expressed in mg Trolox $g^{-1}$ dry weight. The total phenolic compounds of the methanol extracts of aerial parts, as estimated by Folin–Ciocalteu reagent method, was about 460 mg GAE/ g. The phenolic contents in the extracts highly correlate with their antioxidant activity, ($R^2$ = 0.976) confirming that the antioxidant activity of this plant extracts is considerably phenolic contents-dependent.

## INTRODUCTION

Medicinal plants continue being an important resource worldwide, to withstand critical diseases. Overwhelming of the world's population (60-80%) still relies on classical medicines for the curing of current disease (WHO, 2002; Zhang, 2004; Qaralleh et al., 2019). Plants involve various compounds that have active biological substances (Mokbel and Hashinaga, 2006). For instance, phenolic compounds and essential oil play a crucial function as vigorous natural biological factors (Cutter, 2000; Hao et al., 1998; Puupponen-Pimia et al., 2001). The routin use of antibiotic has led to the development of one or more antibiotics resistant infectious bacteria (Sarker et al., 2005; Sufferidini et al., 2004). This issue has resulted on the failure of the treatments to numerous microbial causing infectious disease. Previous investigations pointed out to a number of medicinal plant extracts by constituting a group of potent natural antimicrobial agents (Rayne and Mazza, 2007). Jordan's flora includes more than 2500 wild plant species from 700 genera, and of these, there are around 100 endemic species, 250 rare species, and 125 quite rare species (Tellawi et al., 2001). Although, most of plant species have not been explored chemically or biologically, medicinal plants have been extensively used in traditional medication and continue being a rich source of novel therapeutic agents (Cragg et al., 1997).

Plant family Asteraceae, consists of more than 1500 genera and approximately 2500 species worldwide (Wagstaff and Breitwieser, 2002). The genus *Centaurea* comprises more than 500 species that are common worldwide, particularly the Mediterranean and western Asia area (Mabberley, 1997). Some Asteraceae species have been used in many fields, including nutrition and medicinal industries (Tepe et al., 2006; Tekeli et al., 2011). Many species of Centaurea have long been utilized in conventional medication to treat different diseases, like hemorrhoids, abscess and the common cold (Kargıoglu et al., 2008; Kargıoglu et al., 2010; Sezik et al., 2001). Thus it is worthwhile to elaborate on the antibacterial activities of different species of Centaurea on the case by case basis. Though most of plant species have not been explored chemically or biologically, medicinal plants have been extensively used in traditional medication and continue being a rich source of novel therapeutic agents (Cragg et al., 1997; Al Fraijat et al., 2019). In the present study, we demonstrate that methanolic extracts and essential oil of *C damascene* (Khleifat et al., 2019; Jaafreh et al., 2019), which are used to treat gastritis, and for flavoring hot tea on a small scale in a town south of Jordan (Bsaira). The choice of this species was based on a review of the local folk literature which points out the therapeutic properties that the genus *Centaurea* possess (Alali et al., 2007) and on a preliminary screening study done in our laboratory investigating other medicinal plants for their low cytotoxicity (Qaralleh et al., 2009; Khleifat et al., 2006a; Khleifat et al., 2010; Zeidan et al., 2013; Majali et al., 2015; Althunibat et al., 2016; Al-Asoufi et al., 2017; Qaralleh et al., 2019; ALrawashdeh et al., 2019). The aim of this study was to determine the antibacterial and antioxidant activities as well as toxicity of *Centaurea damascene* that authentically been explored in this work as one of species of *Centaurea* growing in Jordan.

**Chemical Composition of *C. damascena* Essential Oil**

It has been reported that the yield of essential oil of *C. damascena* (table 1). 37 components were characterized on the basis of a typical library search and literature data with selecting only the components showing matches ≥ 90%, which represented about 92.97% of the essential oils from *C. damascena* (Adams, 1995; Flamini et al., 2004; Skaltsa et al., 2003, 2000; Ertugrul et al., 2003; Yayli et al., 2009; Adams, 1995). The main contents of this essential oil consist of 11.45% Fokienol, 8.8% thymol, 8.21% alpha Terpineol, 7.24% Chrysanthemumic acid, 7.13% Terpinen-4-ol and 6.59% Borneol with a high degree of polymorphism in the occurrence of these compounds between the different species of *centaurea* (Fig. 1 and table 2). For example oxygenated monoterpenes was absent or rare in the *C. wageritzii, C. tossiensis* and *C. luschaniana* (Köse et al., 2007). In parallel GC analysis of *C. damascena* showed that Fokienol represent 11.45% whereas *C. wageritzii* showed only 0.50% Fokienol (Köse et al., 2007). However, there is no previous information of investigations on the same species to make a comparison of chemical composition of other regions to show if there is a variation of the major components.

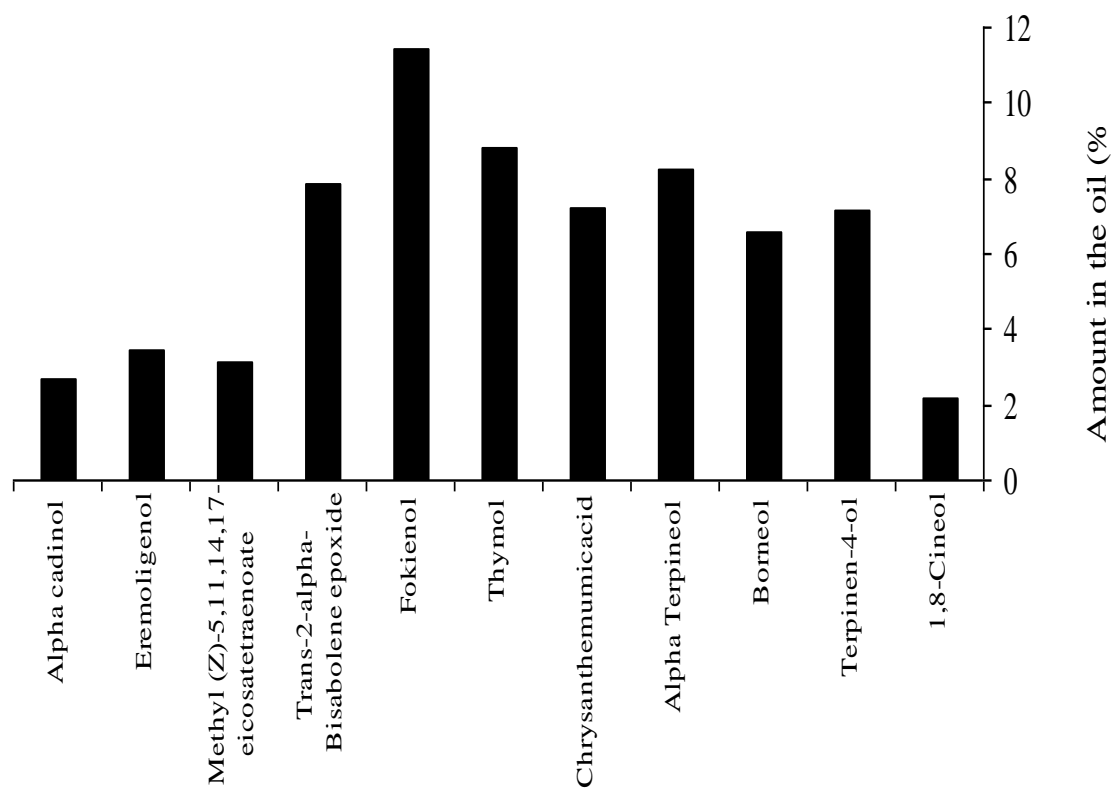

Figure 1. Variation of main constituents (of > 2%) in the essential oil of *C. damascena*.

Table 1. Yield of ethanol crude plant extracts (%)

| Plants | Dry weight (g) | Weight of ethanol extract (g) | Yield of ethanol extract (%) |
|---|---|---|---|
| *C.damascena Bioss* | 50 | 5.72 | 11.44% |

Table 2. The chemical class distribution of the essential oil components of *C. damascena*

| Class compound | Concentration (%) | Number of compounds |
|---|---|---|
| Hydrocarbons | 0.35 | 1 |
| Monoterpenes hydrocarbons | 0.46 | 1 |
| Oxygenated monoterpenes | 46.87 | 14 |
| Sesquiterpene hydrocarbons | 4.41 | 4 |
| Oxygenated sesquiterpenes | 40.89 | 17 |

**Antimicrobial Activities of Essential Oil and Methanolic Extract of *C. damascena***
The regularly used method for evaluation of antibacterial efficiency of publically used medicinal plants can be diffusion as qualitative and dilution methods of the extracts and essential oils as quantitative analyses (Köse et al., 2008). The results show that plant extracts and essential oils have antibacterial activity, while plant extracts did not have that effect as essential oils do on all the tested bacteria. The results gotten from the disc diffusion tests showed that there has been an increasing action of essential oils on the inhibition of bacterial growth on concentration dependent manner. Among the tested microorganisms, *M. lutes* was the most susceptible microorganism against lowest concentration of essential oils. The antimicrobial activity of other *Centaurea* species was focused by many authors (Hajjeh et al., 1999; Güven et al., 2005; Sarker et al., 2005; Yayli et al., 2009; Cansaran et al., 2010 and Tekeli et al., 2011). The essential oils (10 μg/ml) produced the highest inhibition zone of 24 mm that was monitored against *M. lutes*. Oils showed antibacterial activity against seven out of ten tested bacteria. Only *Pseudomonas aeruginosa* and *Klebsiella pneumoniae* were resistant to essential oils of *C. damascene*.

For a more reliable judgment of antimicrobial action (Boğa,et al., 2016), a broth dilution assay was carried out. The MIC of essential oils of *C. damascena* ranged from and 15 ± 1000 μg/mL. The order of MIC values from lowest to highest are *M. luteus* (15μg/ml), *B. subtilis* (50μg/ml), *S. aureus* (50μg/ml), *E. aerogenes* and *S. epidermidis*(123μg/ml), *Salmonella typhi*, *P. mirabilis* and *E. coli* (370μg/ml) and *P. aeruginosa* and *Klebsiella pneumoniae* (1mg/ml). Generally speaking all MIC values for ten bacteria used in this study were in conformity with the results of disc diffusion method. Yayli et al. (2009) obtained a moderate antibacterial activity results on Gram-positive and Gram-negative by the essential oils from *C. appendicigera* and *C. helenioides.* The essential oil of *C. aladagensis* had an antibacterial effect versus seven human pathogenic microorganisms while the essential oils of *C. nicaeensis* and *C. parlatoris* showed low activity on fourteen selected microorganisms (Kose et al., 2007). The antimicrobial characteristic of medicinal plants is really hard to explain.

The methanol extracts of the plants were active against most bacteria being tested with range of inhibition zones (0-19 mm). Among the tested bacteria, *E. aerogenes and S. epidermidis* were the most susceptible bacteria against methanol extracts (19 mm). The extracts exhibited antimicrobial activity against S. aureus, *B. subtilis* and *M. lutes* based on zone of inhibition by 18, 17, and 17 mm, respectively. An important antimicrobial effect of five *Centaurea* species (*C. ptosomipappoides, C. odyssei, C. ptosomipappa, C. amonicola and C. kurdica*) was reported by Güven et al. (2005) on *S. aureus* and *B. cereus*. Conversely, extracts of *C. appendicigera* and *C. helenioides* did not display antimicrobial activity on *E. coli* (Buruk et al., 2006).

The chloroform and ethyl alcohol extracts of *C. cariensis* subsp. *Niveotomentosa* exhibited powerful antibacterial activities on various resistant bacteria, particularly *Staphylococcus* strains (Ugur et al., 2010). The diameter zone of inhibition for tested bacteria by cephixime antibiotic ranged between 9-36 mm. However, for the plant extracts, the average zone of inhibitions observed against these pathogens ranged from 17 to 19 mm. These values fall within the range sensitive and/or intermediate sensitive when compared with control antibiotic. Although, the low values recoded for the plant extracts may be attributed to the fact that the extracts being in crude form, contain very small amounts of bioactive compounds (Olukoya et al., 1993).
For a more reliable estimation of antimicrobial activity, a broth dilution assay was carried out. The lowest MIC value was recorded by *M. lutes* (60 µg mL$^{-1}$) in conformity with the result of disc diffusion method. MIC values of *C. damascene* against *E. coli* and other Gram (-) bacteria were 1100 µg mL$^{-1}$ with an exception of *E. aerogenes* (123 µg mL$^{-1}$) which they are in accordance with results of inhibition zones. The results showed that methanol extracts of the *C. damascena* exerted good antibacterial activity on all the strains being tested at different concentrations.

It is worthy of note that MBC values obtained for the extracts against tested bacteria are higher than MIC, indicating that the extracts are bacteriostatics at lower concentrations and bactericidal at higher concentrations. It noteworthy to mention that even though the MBCs are sometimes 3-fold (≤3300µg/mL) of that the MICs (≤1100µg/mL), the 3.3 mg/mL concentration is still very low as compared with the MICs and MBCs of a lot of medicinal plants in the literature (Tekeli et al., 2011). This suggests that these plant extracts, when used traditionally as antimicrobials inhibit bacterial growth through killing the bacteria and since most of the traditional preparations lack specific concentrations, this may thus account for the use of large quantity of the extracts by traditional medical practitioners for the treatment of their patients. However, the antibacterial activity of most examined *Centaurea* species were shown to be weak or solvent-dependent (Sarker et al., 2005).

However, it has no data previously been reported to discuss any possible varying degree of antibacterial activity by whole *C. damascena* against any type of microorganisms. A study on the biological activities of 12 different species of *Centaurea* was conducted by Tekeli et al., (2011). *B. cereus* was only inhibited by two of the 12 studied *Centaurea* species with 4 mg/ml concentration., *C. calolepis* displayed an action on *Salmonella enteritidis*, while *C. urvillei* subsp. *urvillei* was active against *Escherichia coli* and *Staphylococcus aureus*. In further study on *C. cankiriense* (Cansaran et al., 2010), the MICs of *E. coli* (Abboud et al., 2010; Tarawneh et al., 2009) and *S. aureus* a against ethylacetate and methanol extracts were being as 250 and 62.5 µg/ml, respectively

The antibacterial efficacy in the methanolic extract possibly be due to the presence of tannins, flavonoids, and terpenoids (Zablotowicz et al., 1996; Hemandez et al., 2000; Mamtha et al. 2004). These medically bioactive ingredients perform antimicrobial activity through various mechanisms. Flavonoids, which have been found to be effective antimicrobial substances against a wide array of microorganisms in vitro, are known to be synthesized in response to microbial infection by plants. They have the ability to complex with extracellular and soluble proteins and to complex with bacterial cell walls (Cowan et al., 1999; Khleifat, et al., 2006b; Khleifat, 2006; Khleifat, et al., 2008; Khleifat, et al., 2015).

The saponins have the capacity to rise leakage of metabolites from the cell (Zablotowicz et al., 1996). Tannins cause cell wall synthesis inhibition by forming irreversible complexes with prolene rich protein (Mamtha et al. 2004). Terpenoids cause dissolution of the cell wall of microorganism by undermining the membranous tissue (Hemandez et al., 2000). In this study, *C. damascena* methanol extracts are lacking steroids and this probably why Gram-negative bacteria were less susceptible to *C. damascena* extract than the Gram-positive one. The steroids are known for their antibacterial activity particularly connected with membrane lipids and cause leakage from liposomes (Epand et al., 2007).

**Evaluation of the synergistic effect of antibiotics and essential oil of *C. damascena* on mainly resistant bacteria**
The oil alone from *C. damascena* in the concentration of 20 µg/mL inhibited the bacterial growth of *P. aeruginosa* and *K. pneumonia*. However, a synergetic effect against *E. coli* and *K. pneumonia* was observed when 5 µg/mL oil was combined with some of the antibiotics tested, although some of these antibiotics did not show any activity by themselves such as gentamicin for *E. coli* and vancomycin, ampicillin and chloramphenicol for *K. pneumonia*. Even though a synergistic effect for *P. aeruginosa* with different plant extracts was previously observed (Moon et al., 2011; Nascimento et al., 2000; Rodrigues et al., 2009; Al-Asoufi et al., 2017), no synergetic effect was observed when different antibiotics were combined with 5µg/mL oil to inhibit the growth of *P. aeruginosa*. Probably the concentration factor is critical for synergistic effect since the concentration of 20 µg/mL oil inhibited the bacterial growth of *P. aeruginosa* (Abboud et al., 2009).

In some cases, the mode of action of combinations vary significantly from that of the same antibiotics acting alone. The activity of the antibiotic cefixime was increased by 26% against *K. pneumonia* after combination with the oil (29 mm) since single-handedly cefixime resulted in the inhibition zone of 23 mm, showing that this oil at 5 µg/mL influences the activity of the antibiotic and may be used as a potential adjuvant in the antibiotic cure of pneumonia caused by *K. pneumoniae*.

**Antioxidant Activity of of *C. damascena* Essential Oil**

DPPH is normally employed as a reagent to estimate free radical scavenging capacity of antioxidants (Zengin et al., 2010). Free radical scavenging capacity of the *C. damascna* methanol extract was calculated by DPPH test. IC50 value is expressed by which the functional concentration of 50% DPPH radicals were scavenged and was determined from the graph plotting inhibition percentage versus extract concentration. $IC_{50}$ of methanol extract was detected by using DPPH assay and expressed as µg/ml. IC50 estimate is inversely correlated to antioxidant capacity of extracts. In the present study, *C. damascena* extract exerted better antioxidant efficiency than that of previously reported *Centaurea* species extract with an IC50 value of 17.08 µg /ml. The extract exhibited concentration-dependent radical scavenging activity,

As far as we know from literature survey, the *Centaurea damascena* have shown better activities than other *Centaurea* species such as *C. patula, C. pulchella,* and *C. tchihatchfeii* (Zengin et al., 2010), *C. mucronifera* (Tepe et al., 2006), *C. huber-morathii* (Sarker et al., 2005) and *C. centaurium* (Conforti et al., 2008) by the same DPPH assay. Antioxidant capacities obtained by the FRAP was 51.9 and expressed in mg Trolox $g^{-1}$ dry weight. The two methods (FRAP and DPPH) approximately showed similar results for *C. damascena* methanol extract since the two methods, the assay results were generally similar for different plant models because these methods are based on electron transfer mechanism (OZGEN et al., 2006)

The total phenolic compounds of the methanol extracts of aerial parts, as estimated by Folin–Ciocalteu reagent method, was about 460 mg GAE/ g. The highest level of phenolics was found in *C. damascena* as compared with previously reported other species of *Centuarea* (Sarker et al., 2005; Tepe et al., 2006; Karamenderes et al., 2007; Alali et al., 2007; Conforti et al., 2008; Dudonné et al., 2009; Zengin et al., 2010). The results of total phenolic compounds suggested that the phenolic contens contributed significantly to the antioxidant ability of the *C. damascena extract*. It has been reported in the literature that phenolic contents have strong antioxidant capacities and these phenolics have antioxidant activity mainly as a result of their redox properties which enable them to act as reducing agents, singlet oxygen quenchers, hydrogen donators, and metal chelator (Baratto et al., 2003; Alali et al., 2007; Zengin et al., 2010). However, it was found that methanolic plant extracts are the most effective scavenger of DPPH radical as compared with other solvents (Miliauskasa et al., 2004). Hence, it was suggested that methanol is more efficient solvent for cell walls and seeds degradation, that have nonpolar nature causing the release of polyphenols from cells. It can be observed that the phenolic contents in the extracts highly correlate with their antioxidant activity, (R2 = 0.97), confirming that phenolic compounds contribute significantly to the antioxidant activity of these plant extracts (Fig. 2). The enormous disparity in Centaurea species antioxidant activity could result from variations in total phenolic contents (Al-Mustafa, & Al-Thunibat, 2008). Such observation agreed with several previous findings (Sarker et al., 2005; Tepe et al., 2006; Guven et al., 2005; Karamenderes et al., 2007; Alali et al., 2007; Conforti et al., 2008; Zengin et al., 2010; Karahan et al., 2016).

Moreover, phenolic compounds have various reactions to Folin-Ciocalteau assay (Sun et al., 2002). The molar response of this method is roughly proportional to the number of phenolic hydroxyl groups in a given substrate, but the reducing capacity is enhanced when two phenolic hydroxyl groups are oriented in ortho or para-position. Since these structural features of phenolic compounds are responsible for antioxidant activity (Katalinic et al., 2006). Thus, polyphenols measurements in extracts may be related to their antioxidant activities.DPPH and FRAP methods have been used by many researchers to evaluate the free radical scavenging activity of antioxidant molecules and plant extracts. DPPH does generate strongly colored solutions with methanol which is eliminated in presence of antioxidants. The data obtained from the two radical scavenging methods suggest high accuracy and constancy between them.

Using two methods, some plants showed large difference in their TEAC values, whereas others showed little differences. This may be due to variation in types of phenolic compounds, that differ significantly in their reactivity towards DPPH (Katalinic et al., 2006; Benzie and Strain, 1996). Furthermore, the affinities of different Centaurea species toward the two above radicals were sometimes significantly altered due to variation of DPPH solubility in aqueous medium.

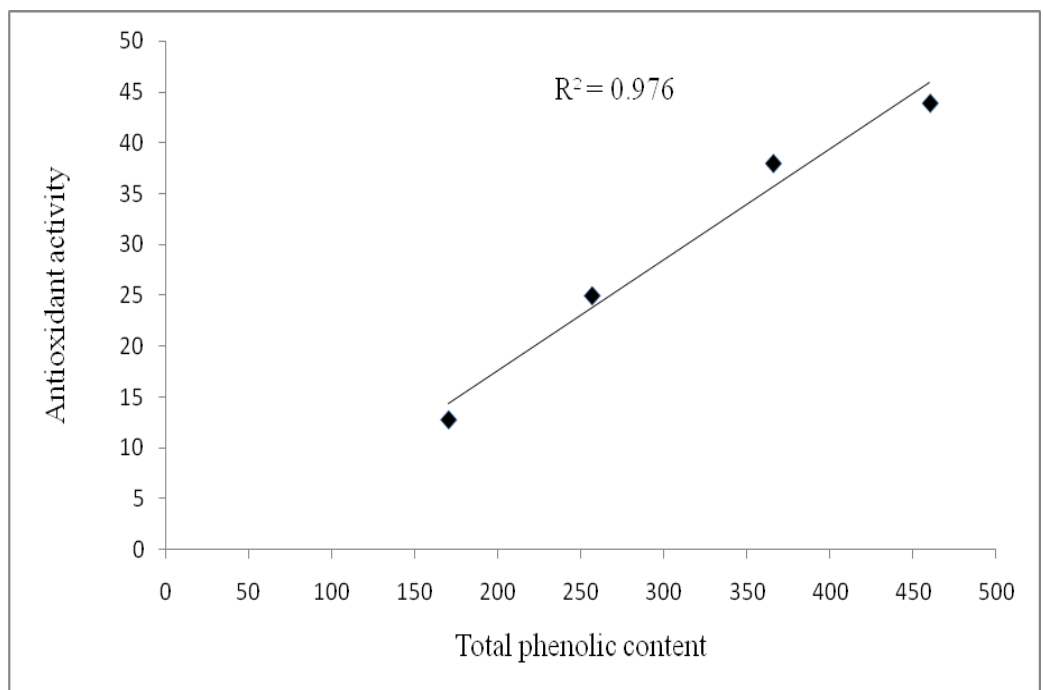

Figure 2. Linear correlation between the total phenolic content and antioxidant activity TEAC for methanolic extracts: correlation coefficient r = 0.822, coefficient of determination r$^2$ = 0.976.